\documentclass[12pt,a4paper]{article}

\usepackage[british]{babel}

\usepackage[a4paper,top=2cm,bottom=2cm,left=2.5cm,right=2.5cm,marginparwidth=1.75cm]{geometry}

\usepackage[style=nature, backend=biber]{biblatex} 
\DeclareLanguageMapping{british}{british-apa} 
\DeclareFieldFormat[article]{volume}{\textbf{#1}} 

\usepackage{amsmath}
\usepackage{graphicx}
\usepackage[colorlinks=true, allcolors=blue]{hyperref}
\usepackage{orcidlink}
\usepackage[title]{appendix}
\usepackage{mathrsfs}
\usepackage{amsfonts}
\usepackage{booktabs} 
\usepackage{caption}  
\usepackage{threeparttable} 
\usepackage{algorithm}
\usepackage{algorithmicx}
\usepackage{algpseudocode}
\usepackage{listings}
\usepackage{enumitem}
\usepackage{chngcntr}
\usepackage{booktabs}
\usepackage{lipsum}
\usepackage{subcaption}
\usepackage{authblk}
\usepackage[T1]{fontenc}    
\usepackage{csquotes}       
\usepackage{diagbox}
\usepackage{xcolor}
\usepackage{pdfpages}
\usepackage[normalem]{ulem}
\newcommand\redsout{\bgroup\markoverwith{\textcolor{red}{\rule[0.5ex]{2pt}{0.75pt}}}\ULon}

\usepackage{setspace}
\usepackage{titlesec}

\titleformat{\section}
  {\normalfont\Large\bfseries}{}{0pt}{}
  
\usepackage{float}   
\usepackage{caption} 


\newcommand{\revnew}[1]{\textcolor{black}{#1}}
\newcommand{\revnewtwo}[1]{\textcolor{black}{#1}} 

\title{Advancing Scientific Discovery and Complex Optimization through Distributed Quantum Neural Networks
}

\author{
	Seongmin Kim and In-Saeng Suh \\
        \small National Center for Computational Sciences, \\
        \small Oak Ridge National Laboratory, Oak Ridge, Tennessee 37830, United States. \\
	\small Corresponding author. Email: kims@ornl.gov, suhi@ornl.gov
}

\date{}
\begin{document}
\maketitle
\date{}

\begin{abstract}
Optimization problems are critical across various domains, yet existing quantum algorithms, despite their great potential, struggle with scalability and accuracy due to excessive reliance on entanglement. To address these limitations, we propose variational quantum optimization algorithm (VQOA), which employs an ansatz based solely on quantum superposition, where single-qubit rotation gates function analogously to neurons in classical deep neural networks. This ansatz, which can be regarded as quantum neural networks, significantly reduces circuit complexity, enhances noise robustness, mitigates Barren Plateau issues, and enables efficient partitioning for highly complex, large-scale optimization. Furthermore, we introduce distributed VQOA (DVQOA), which integrates high-performance computing with quantum computing to achieve superior performance. These features enable a significant acceleration of material optimization tasks (e.g., metamaterial design), achieving more than 50$\times$ speedup compared to state-of-the-art optimization algorithms. Beyond material design, DVQOA efficiently solves quantum chemistry problems and \textit{N}-ary $(N \geq 2)$ optimization problems involving higher-order interactions, outperforming classical deep neural networks. These advantages establish DVQOA as a highly promising and versatile solver for real-world problems, demonstrating the practical benefits of the quantum-classical approach. 
\end{abstract}

\textbf{Keywords}: quantum neural networks, distributed computing, N-ary optimization, higher-order interactions, materials optimization

\section{Introduction}
Quantum computing leverages the principles of quantum mechanics to process information in ways fundamentally different from classical computing \cite{ladd2010quantum, herr2017optimization}. These unique properties provide a theoretical advantage for solving optimization problems, especially those formulated as energy minimization tasks on complex solution landscapes, where the goal is to find the ground state of a cost Hamiltonian \cite{o2007optical, dupont2023quantum}. Quantum algorithms excel in this domain by efficiently navigating the solution space, exploring multiple possibilities simultaneously, which classical computing cannot achieve \cite{divincenzo1995quantum, grover2000synthesis}. This makes quantum computing particularly promising for combinatorial optimization problems, known as NP-hard \cite{schuetz2022combinatorial}. Variational quantum algorithms (VQAs), based on the variational principle \cite{cerezo2021variational} (Supplementary Text), have emerged as one of the most promising approaches to solving such problems by leveraging quantum computing (QC) in conjunction with classical computing \cite{wang2024comprehensive, leymann2020bitter}. VQAs combine parameterized quantum circuits (PQCs) with classical optimization: PQCs encode potential solutions while classical optimizers iteratively adjust the parameters to minimize a cost function \cite{abbas2024challenges, cerezo2021variational}. 

VQAs, such as quantum approximate optimization algorithm (QAOA), are primarily designed to handle binary optimization tasks, aligning with the binary nature of qubits \cite{gambella2020multiblock, shaydulin2024evidence, larkin2022evaluation}. This inherently restricts their utility for \textit{N}-ary optimization problems (e.g., ternary or quaternary tasks), where each variable assumes multiple states. This limits their applicability to broader optimization tasks that require multi-state variables. Moreover, VQAs face challenges when dealing with higher-order interactions, which involve interactions among multi-variables. Encoding these interactions requires numerous two-qubit entangling gates for quantum circuits, dramatically increasing circuit depth and computational complexity due to the combinatorial explosion in the number of two-qubit gates required as problem size increases \cite{pelofske2024short, pelofske2024scaling}. Since two-qubit gates are prone to error and introduce computational overhead, this limits VQAs to efficiently solving higher-order problems \cite{huang2019fidelity, yamamoto2023error, corcoles2015demonstration}. Consequently, for \textit{N}-ary optimization problems involving higher-order (\textit{k}$^{th}$-order) interactions, problem complexity scales as \textit{O}(\textit{N$^{n}$n$^{k}$}) with \textit{n} variables (Supplementary Text), making these problems extremely challenging. Therefore, a highly efficient ansatz with reduced algorithmic complexity is required to tackle such challenges.

Furthermore, VQAs may fail to reach the ground truth, especially for complex problems \cite{amaro2022case, azad2022solving, cerezo2021variational}. This limitation can be mitigated by employing high-performance computing (HPC), which provides the computational capacity to explore broader solution spaces through distributed computing \cite{farooqi2023exploring}. Although the integration of HPC and QC has great potential for addressing real-world problems \cite{beck2024integrating}, research on their integration to address these problems has been limited. This highlights the need for further investigation into how HPC can be optimally leveraged to enhance the performance of quantum algorithms, especially in improving solution quality by leveraging the strengths of both HPC and QC resources.


In this work, we present a variational quantum optimization algorithm (VQOA) designed to efficiently solve various real-world problems. VQOA features a highly efficient ansatz that relies solely on quantum superposition within a many-qubit (MQ) system (Fig. 1a), significantly reducing algorithmic complexity and computational cost. The MQ system refers to a quantum circuit system composed of numerous unentangled single-qubit rotation gates acting independently across many qubits, enabling a quantum-parallel approach. In this framework, the single-qubit rotation gates act like neurons in classical neural networks, positioning the MQ system as a quantum analog of a neural network ansatz. This MQ system enables seamless circuit partitioning without losing any quantum information, allowing for large-scale quantum simulations. Furthermore, this algorithm achieves superior performance by utilizing HPC-QC integrated systems (termed Distributed VQOA; ``DVQOA"), demonstrating its capability to practically utilize quantum-centric supercomputing architectures. With these unique capabilities, DVQOA effectively addresses a wide range of real-world problems, including \textit{N}-ary (\textit{N}~$\geq$~2) problems that involve higher-order (\textit{k}~$\geq$~2) interactions, highlighting the practical \revnewtwo{benefits} of the quantum-classical approach.

\section*{Results}
\subsection*{Efficient ansatz and circuit partitioning}

The efficient ansatz is the core of VQOA, enabling remarkable scalability and adaptability for various optimization challenges. Conventional VQAs, such as QAOA, are designed to approximate the optimal solution of combinatorial optimization problems, but often rely on complex quantum circuits with numerous two-qubit entangling gates, which increase circuit depth and introduce significant errors \cite{pelofske2024short, pelofske2023high}. The primary difference between QAOA and VQOA lies in their circuit design and parameter optimization. QAOA employs a variational ansatz consisting of alternating layers for a problem-specific cost Hamiltonian ($H_C$) and a mixing Hamiltonian ($H_M$), with each layer having two trainable parameters: one for the cost Hamiltonian and one for the mixing Hamiltonian. However, as problem size increases, QAOA faces significant challenges due to heavily entangled two-qubit gates (such as CNOT gates), leading to performance degradation caused by accumulated gate errors, decoherence, and hardware connectivity constraints. While these challenges may be mitigated in the future with the development of fault-tolerant quantum computers, their resolution remains contingent on advancements in quantum error correction and scalable quantum architectures.

In contrast, VQOA employs the MQ system (i.e., an ansatz that entirely eliminates two-qubit gates, relying solely on single-qubit rotation gates), significantly reducing circuit complexity and minimizing noise effects (Fig. 1a). Each rotation gate is associated with an independent parameter, which is optimized using a classical optimizer. Similar to classical neural networks, these parameters are updated iteratively to minimize the cost functions. Through this process, the parameterized rotation gates naturally capture higher-order interactions among variables, enabling VQOA to execute considerably faster and more accurately than conventional VQAs. In VQOA's ansatz, \textit{R$_x$} and \textit{R$_y$} gates are the effective gates to determine computational states after measurements on the \textit{z}-basis (Fig. S1) \cite{xu2023classical}. However, using both \textit{R$_x$} and \textit{R$_y$} gates unnecessarily increases the number of trainable parameters, potentially degrading performance. By exclusively using \textit{R$_y$} gates, the circuit avoids unnecessary complexity while maintaining effectiveness. This simplified ansatz ensures that the interactions among variables can be effectively captured through parameter optimization without using entangling gates. 

\revnew{For quantum hardware experiments, the maximum number of optimization iterations is constrained to 200 to balance convergence requirements with practical quantum computing costs. Empirically, small and moderate-size problems typically converge within this range, and larger problems exhibit clear convergence trends that enable fair comparison between hardware and simulator performance.} While the approximation ratios achieved on the \texttt{IBM-Strasbourg} quantum device and noiseless simulator may not reach the ground truth due to the constrained iterations, their overall performance remains comparable, demonstrating the algorithm's robustness against errors introduced by hardware imperfections (Fig. 1b). This high performance on hardware is attributed to the exclusive use of error-robust single-qubit gates in the ansatz, highlighting the practicality of implementing this quantum algorithm on real quantum devices. Notably, quantum algorithm execution scales significantly differently between hardware and simulator. \revnewtwo{On quantum hardware}, execution time remains largely independent of circuit width (i.e., problem size, \textit{n}), exhibiting \revnewtwo{nearly} constant time complexity. This demonstrates the suitability of quantum devices for tackling large-scale problems. In contrast, on the simulator, execution time grows exponentially with circuit width, as the simulator should classically track all possible quantum states and perform quantum operations accordingly (Fig. 1c, Supplementary Text)\cite{guerreschi2020intel, efthymiou2021qibo}. 

Despite the demonstrated speedup on hardware, limited access to quantum device resources necessitates continued reliance on quantum simulators for research and development of quantum algorithms. A key strength of our approach lies in the absence of two-qubit entangling gates, which enables seamless partitioning of quantum circuits without any loss of quantum information. This feature allows the circuit to be divided into smaller, independent segments, significantly enhancing scalability for large-scale quantum simulations. Partitioning the circuit greatly reduces the overall computational effort required to solve large problems while maintaining high approximation ratios. For example, solving problems (\textit{n} = 30) without partitioning would require substantial computational time using a simulator. However, by partitioning the circuit into smaller pieces, the time to solution is exponentially reduced without compromising solution quality due to the significantly reduced memory demand for simulations (Fig. S2). This indicates that single-qubit simulation is feasible, where each qubit is simulated independently, which results in a linear increase in time to solution with problem size (Fig. 1c). However, for moderately wide circuits, simulating the entire circuit without partitioning can be more efficient, highlighting the importance of optimizing the number of segments for efficient wide-circuit simulations (Fig. 1c). Furthermore, considering the time scaling of VQOA on quantum hardware and simulators, the results show that quantum algorithms may achieve computational speedup on quantum hardware over simulators beyond a certain problem size, even after circuit partitioning. \revnewtwo{We note that the VQOA ansatz belongs to the class of separable variational circuits \cite{misra2023mean}. Consequently, interactions among variables are not represented through quantum entanglement but are instead captured through classical optimization of the circuit parameters guided by the cost function. This design makes VQOA particularly suitable for classical optimization tasks, such as combinatorial optimization, where the objective is to minimize a cost function defined over classical variables. Hence, VQOA remains effective for optimization landscapes and offer practical benefits, including reduced circuit depth, improved noise robustness, and scalability on near-term devices.}

There are two key hyperparameters in the ansatz: the number of layers (\textit{m}) and repeats (\textit{t}). The hyperparameter \textit{m} controls the number of trainable parameters in the circuit, while \textit{t} repeats the circuit structure to enhance its expressive power, as illustrated in Fig. 1a. Excessively high values of \textit{m} can lead to lower approximation ratios due to an unnecessary increase in the number of trainable parameters, but this issue can be efficiently mitigated by increasing \textit{t} while keeping \textit{m} small. Repeated rotation gates (\textit{t}) enhance circuit complexity to explain complex solution spaces without excessively increasing the number of trainable parameters. VQAs often encounter challenges in identifying ground truth due to the Barren Plateaus phenomenon, where the gradient of the cost function diminishes exponentially, hindering effective optimization. This issue can be addressed by employing a reasonable number of trainable parameters with fewer entangling gates \cite{mcclean2018barren, bravyi2020obstacles, sack2022avoiding}. \revnew{Because VQOA eliminates all two-qubit entangling gates, it fundamentally reduces parameter scaling compared to traditional VQAs. Algorithms such as QAOA require a combinatorial number of entangling operations to encode pairwise or higher-order interactions, leading to substantial circuit depth and parameter growth. In contrast, VQOA captures interactions implicitly through optimized single-qubit rotations, avoiding the significant parameter and depth growth. Consequently, by reducing the number of trainable parameters without entanglement, our approach effectively suppresses the Barren Plateaus problem, thereby resulting in improved approximation ratios (Fig. 1d,e,f).} Here, increasing both \textit{m} and \textit{t} results in a linear increase in time to solution, leading to an algorithmic complexity of \textit{O}(\textit{nmt}) (Fig. 1g,h,i). To balance computational efficiency and solution quality, specific values of \textit{m} and \textit{t} are selected. For smaller problems (\textit{n}~$\leq$~20), they are set to 3, whereas they are set to 7 for larger problems (\textit{n}~$>$~20). These hyperparameters provide robust performance across a wide range of problem sizes while maintaining acceptable computational costs.

\subsection*{HPC-QC integration}

In optimization, it is crucial to explore diverse regions of a solution space to find the global optimum. Integrating HPC with QC to leverage multi-cores/nodes can significantly enhance the solution quality by enabling the distributed execution of quantum algorithms with varying initial parameters. This ensures comprehensive exploration of the solution space, increasing the chances of identifying the global optimum. Although increasing the number of cores (\textit{c}) can introduce communication overhead, its impact on time to solution is modest. For instance, utilizing 500 cores results in only a $\sim$41\% increase in time to solution compared to using 10 cores, primarily due to variations in convergence speeds across executions (Fig. S3). These results highlight the practicality of distributed execution for achieving high-quality solutions while maintaining computational efficiency. Therefore, to maximize the solution quality, ideally approaching the ground truth, DVQOA employs 500 cores (with 10 compute nodes) for further studies. 

Brute force search guarantees the identification of the ground truth \cite{giuffrida2022engineering}. However, its runtime for quadratic unconstrained binary optimization (QUBO) problems increases exponentially with \textit{n}, making it computationally challenging for large-scale problems. While utilizing multiple cores (\textit{c}) can reduce computation time linearly, the overall time complexity remains exponential at \textit{O}(2$^{n}$/\textit{c}) (Fig. S4). In contrast, DVQOA achieves significantly higher efficiency with a linear time complexity of \textit{O}(\textit{nmt}), allowing it to find the ground truth substantially faster than brute force methods (Fig. 2a). However, when the number of qubits does not divide evenly into the partition numbers, the approximation ratio may degrade. This issue can be simply mitigated by adjusting the number of partitions to ensure even distribution, which has minimal impact on the time (Fig. S5). 

QAOA is widely used for tackling QUBO problems \cite{cerezo2021variational, turati2022feature}. However, its reliance on many two-qubit operators in the cost layers leads to deep circuits, making QAOA inefficient in current quantum computing systems (quantum hardware and simulators) \cite{pelofske2024short, pelofske2023high}. Consequently, QAOA exhibits low approximation ratios and long time to solutions (Fig. 2b,c) \cite{kim2024distributed}. On the other hand, DVQOA features significantly shallow and efficient circuits, thus it can achieve much shorter time with higher approximation ratios. Moreover, DVQOA demonstrates robust scalability, effectively solving larger problems that QAOA struggles with (\textit{n} = 32, 36, and 40; Fig. S5). Note that DVQOA can be implemented on quantum-centric supercomputing architectures, utilizing either a single quantum device with qubit clusters or multiple quantum devices for distributed processing (Fig. S6).

\subsection*{Efficiency of DVQOA for real-world problems}

DVQOA can be used for addressing various problems, simply redefining its cost function with the same ansatz (Fig. 3a). To demonstrate this, DVQOA is applied to solve Max-Cut problems, typical examples of binary combinatorial optimization \cite{commander2009maximum}. Their problem complexity grows exponentially with increasing number of nodes (Fig. S7) \cite{schuetz2022combinatorial}. Despite their complexity, DVQOA reliably converges to the ground truth, achieving approximation ratios of 1 (Fig. 3b). Similar to its performance on QUBO problems (Fig. S5), QAOA struggles with modest-scale Max-Cut problems (e.g., \textit{n} = 30), yielding low approximation ratios and long time to solutions (Fig. S8). On the other hand, our quantum algorithm demonstrates remarkable scalability, solving problems with up to 1,000 nodes, which are far beyond the reach of brute force search and QAOA. For such large-scale Max-Cut problems, hybrid quantum annealing (HQA), one of the best solvers for these problems (Table S1) \cite{schuetz2022combinatorial, benlic2013breakout, kochenberger2013solving}, is used as a reference for comparison. Although this experiment does not include distributed executions, our VQOA consistently achieves high approximation ratios ($>$~0.93) and maintains short times to solution ($<$~1,000~s) even without hyperparameter optimization for such scales (Fig. S9). Additionally, DVQOA demonstrates its versatility to handle other optimization problems, e.g., traveling salesman problems (TSP). It identifies the optimal routes with (near)-optimal distances, outperforming a classical solver (simulated annealing; SA) and achieving performance on par with a leading quantum solver (HQA) for these problems (Fig. 3c, Fig. S10, and Table S2). 

It should be noted that DVQOA applies not only to optimization tasks but also to minimum eigenvalue calculation through cost function evaluation, e.g., computing the minimum eigenvalues of a Hamiltonian representing a molecule. Variational quantum eigensolver (VQE) with a two-local ansatz has been used for quantum chemistry to identify the minimum energy state of molecules \cite{xu2024truncation, tran2024variational}. While VQE with the two-local ansatz achieves high accuracy for simple molecular structures, such as hydrogen, its performance deteriorates for more complex systems, indicated by lower approximation ratios (see Methods for more details). In contrast, our algorithm, which exclusively employs single-qubit operators in the ansatz, consistently identifies low-energy states (Fig. 3d). These results clearly show DVQOA’s high performance not only in optimization but also in eigenvalue computations, reinforcing its broad applicability. \revnewtwo{However, because DVQOA employs an ansatz composed solely of single-qubit operations without entangling gates, its representational capacity may be limited for systems where strong quantum correlations are essential. In particular, accurately describing the ground states of correlated molecular systems generally requires entangled wavefunctions \cite{bauer2020quantum, kim2026harnessing}. Consequently, while the framework can identify low-energy configurations for small or weakly correlated molecular systems, it is not expected to reach the exact ground state for strongly correlated molecules or to systematically achieve chemical accuracy (approximately 1.6 mHa). Incorporating entangling operations into the ansatz could enhance the expressive power of the framework and improve its applicability to more challenging quantum chemistry problems. Exploring such extensions is beyond the scope of the present study.}

Importantly, DVQOA's exceptional efficiency extends to material optimization tasks, such as layered photonic structures designed for energy-saving windows (Fig. S11, see Methods for more details)  \cite{kim2021visibly, kim2022high, so2024radiative}. Fig. 3e shows that DVQOA achieves convergence well, resulting in a highly optimized structure that outperforms the best-known result in the field. This optimized structure exhibits improved performance indicated by a lower figure-of-merit (FOM) \cite{kim2022high}. To further showcase DVQOA's potential, it is employed to design metamaterial optical diodes, which aim to enable unidirectional light transmission (Fig. S12, see Methods for more details) \cite{kim2024quantum, xu2024quantum}. The results in Fig. 3f reveal that DVQOA not only converges effectively but also identifies a structure with better performance than the best-known material \cite{kim2024quantum}. 

These material optimization tasks are particularly challenging due to the exponential growth of the optimization space arising from the combinatorial explosion of possible configurations \cite{kitai2020designing, kim2022high, kim2024quantum}. Despite the inherent exponential problem complexity, DVQOA exhibits remarkable efficiency, requiring less than one second per iteration and completing the optimization of 40-qubit problems within $\sim$25 minutes (Fig. 3g). This presents a significant acceleration of DVQOA ($>$~50$\times$) compared to state-of-the-art quantum computing-assisted active learning algorithms designed for material optimization, which require $\sim$1,342 minutes \cite{kitai2020designing, kim2022high, kim2024quantum, kim2024distributed}. DVQOA achieves this speed by avoiding the need for surrogate modeling, a common approach in active learning algorithms to approximate optimization spaces \cite{lookman2019active, jablonka2021bias}. Here, surrogate modeling introduces additional computational overhead, often requiring tens of seconds through machine learning techniques \cite{lookman2019active, jablonka2021bias, kim2024review}. In contrast, DVQOA inherently functions like a quantum active learning algorithm by directly leveraging its ansatz as a surrogate model to minimize the cost function, making it considerably faster than other active learning algorithms. These results highlight DVQOA’s groundbreaking efficiency, particularly in material science, by significantly reducing computational time while achieving higher-quality outcomes. Furthermore, these demonstrate that DVQOA is not limited to QUBO-type problems but can effectively solve a broad range of real-world problems.

\subsection*{Higher-order (\textit{k}$^{th}$) \textit{N}-ary optimization challenges}

DVQOA can inherently capture higher-order interactions among multi-variables, enabling it to capture more complex relationships during optimization. Table 1 proves DVQOA’s ability to accurately solve problems involving higher-order interactions (\textit{k} = 3, 4, and 5). This capability ensures the identification of superior optimization results with lower FOM values in material optimization tasks, compared to the active learning algorithms (Fig. 3e,f). Quantum solvers such as QA, HQA or QAOA, used in the active learning algorithms, are generally limited to handling 2$^{nd}$-order problems \cite{kim2022high, kim2024quantum, kim2024distributed}. This limitation prevents them from capturing the complexities of higher-order interactions \cite{kim2022high, kim2024quantum}, leading to suboptimal outcomes. DVQOA’s enhanced capability overcomes these limitations, enabling it to achieve significantly better results. Furthermore, a performance comparison between DVQOA and a classical machine learning algorithm (distributed execution of deep neural networks, DDNN), presented in Table 1 and Table S3, reveals that DVQOA outperforms the DDNN, achieving higher approximation ratios and significantly shorter time to solutions, particularly for more challenging problems (Supplementary Text). \revnewtwo{These results highlight practical benefits of the quantum approach over classical algorithms.}

The inherent binary nature of qubits \cite{gambella2020multiblock, shaydulin2024evidence}, which represent states 0 and 1, poses challenges for conventional VQAs in modeling and optimizing \textit{N}-ary problems. DVQOA overcomes this limitation by assigning multiple states on the Bloch Sphere to represent different labels \cite{herrmann2023first}. This approach maps a state vector to a specific label (Fig. 4a), allowing each qubit to represent one among multiple states ($\geq$ 2), thereby enabling DVQOA to effectively solve \textit{N}-ary optimization challenges. These \textit{N}-ary problems, such as ternary or quaternary optimization, present exponentially larger solution spaces (3$^{n}$ or 4$^{n}$) compared to binary optimization problems (2$^{n}$), making them significantly more challenging to solve. Additionally, the inclusion of higher-order interactions further increases the complexity, resulting in an exponential problem complexity of \textit{O}(\textit{$N^{n}n^{k}$}) (Supplementary Text). DVQOA can successfully solve such highly intricate problems with a linear complexity \textit{O}(\textit{nmt}), reliably identifying the ground truth (Table S4). For these complex scenarios, hyperparameter tuning may be required to efficiently capture complex relationships for the identification of the ground truth (Table S4,5). Table 2 presents DVQOA’s exceptional capability to accurately solve \textit{N}-ary optimization problems (binary, ternary, quaternary, and quinary) involving 2$^{nd}$ and $3^{rd}$-order interactions within a minute. 

To further demonstrate DVQOA’s capability in solving \textit{N}-ary optimization problems in material science, layered photonic structures are optimized where each layer can take one of three material candidates (ternary problem). While conventional methods require additional qubits and constraints to represent multiple states \cite{kim2022high}, DVQOA’s state-vector assignment on the Bloch sphere efficiently eliminates this requirement \cite{herrmann2023first}, leading to superior optimization results (Fig. 4b, Supplementary Text). Moreover, increasing \textit{m} and \textit{t} allows more parameter adjustments, resulting in further optimized structures with lower FOMs. As shown in Fig. S13, the optimized photonic structure exhibits improved optical characteristics. Fig. 4c presents a linear time complexity of DVQOA, demonstrating its scalability and efficiency in solving real-world multi-state problems. This exceptional performance of DVQOA makes it a unique and revolutionary tool to handle various optimization tasks across diverse domains.

\revnew{We note that although the proposed DVQOA is developed primarily for discrete optimization, it can be naturally extended to continuous problems through discretization. Continuous variables may be mapped onto a finely discretized set of values, allowing each qubit (or each multi-state variable encoded through Bloch-sphere representation) to correspond to a discrete value. By increasing the discretization resolution, the continuous domain can be approximated with arbitrarily high fidelity. This approach aligns well with DVQOA's ability to handle \textit{N}-ary variables, providing a flexible and scalable pathway for solving discrete, continuous, and mixed-integer optimization tasks.}

\section*{Discussion}
We have proposed a highly efficient ansatz that relies exclusively on single-qubit gates, capturing interactions among multi-variables through parameter optimization. This ansatz significantly reduces circuit complexity while preserving the ability to capture higher-order interactions through parameter optimization, making it an effective quantum analog of classical DNNs. This approach allows quantum circuits to be seamlessly partitioned without loss of quantum information, enabling large-scale quantum simulations. The algorithm’s performance is significantly enhanced with high accuracy through distributed execution with diverse initial parameters on HPC-QC systems. The versatility to tackle various real-world problems, including quantum chemistry calculations, is another key advantage of DVQOA. Furthermore, DVQOA functions like quantum active learning, significantly accelerating (more than 50$\times$) material optimization tasks and achieving superior results compared to state-of-the-art optimization algorithms. Remarkably, the algorithm extends beyond binary optimization by enabling multi-state optimization, where each qubit can represent multiple states on the Bloch sphere. This highlights DVQOA’s capability for effectively addressing higher-order \textit{N}-ary optimization challenges. Importantly, our results demonstrate a practical \revnewtwo{benefit} of utilizing the quantum approach in optimization, with quantum hardware exhibiting \revnewtwo{nearly} constant complexity and the quantum-classical algorithm outperforming well-established classical algorithms. These findings establish DVQOA as a powerful and versatile tool for solving challenging problems utilizing current QC technologies. Therefore, we expect that DVQOA will play a critical role in optimization and eigenvalue calculation across various fields. Moreover, this work lays the groundwork for leveraging near-term quantum devices in solving real-world challenges effectively.

\section{Methods}

\subsection*{Computational experiments}

We use qiskit (version 0.41.0) for developing and executing the quantum algorithm, while the newer version (1.2.4) is employed for the execution of the quantum algorithm on quantum hardware. For quantum simulation, we utilize qiskit-aer (version 0.11.2) with the 'automatic' or 'statevector' method. For classical optimization, we employ the 'scipy.optimize.minimize' (SciPy version 1.2.1) with a gradient-free optimizer 'COBYLA' to minimize the cost function. Parameters in rotation gates in our ansatz are randomly initialized within the range [-2$\pi$, 2$\pi$], and these parameters are iteratively adjusted through classical optimization. 

To implement our quantum algorithm on quantum hardware and compare its performance with the simulator, the maximum number of iterations was set to 200, considering hardware access limitations and ensuring practical execution times. The hyperparameters (number of layers \textit{m} and repeats \textit{t}) are set to 3, and the number of partitions is set to 1 for this analysis. Specifically, we utilize the \texttt{IBM-Nazca} and \texttt{IBM-Strasbourg} quantum devices, which feature 127 qubits, to evaluate and validate the algorithm's performance on real quantum computing devices. For this study, a MacBook Pro, equipped with an Apple M2 Max processor and 32 GB of memory, is employed for submitting quantum circuits and processing the classical components of the algorithm without leveraging distributed execution.  

Despite the observed \revnewtwo{benefit} on quantum hardware, quantum simulators are used for most studies due to limited access to quantum devices. \revnewtwo{Therefore, results are obtained from quantum simulators unless explicitly stated otherwise for quantum hardware executions.} To maintain the efficiency of our quantum algorithm on the simulator, circuit widths are capped at 10 qubits (Fig. S2). The optimization process terminates upon achieving full convergence (by default tolerance) or reaching a predefined maximum iteration limit (5,000 in this study). For small-scale problems, convergence is usually achieved before reaching the maximum iterations. However, for large-scale problems (\textit{n}~$\geq$~100), convergence tends to be less stable, leading to unnecessarily extended optimization runs, thus increasing computational costs. To address this issue, an additional stopping criterion is introduced: if the cost value changes by less than 0.05\% over the 500 consecutive iterations, the optimization process is terminated. This dual-criterion approach ensures computational efficiency while maintaining accuracy across diverse problem scales. For scalability studies, 100- to 1,000-qubit Max-Cut problems with five different instances are addressed without distributed execution using the MacBook Pro. In addition, hyperparameters (\textit{m} and \textit{t}) are set to 3.

\subsection*{Integrating high-performance computing (HPC) with quantum computing (QC)}

Two HPC systems are utilized for the integration with QC: “Frontier” and “Defiant”, both located at the Oak Ridge Leadership Computing Facility. While these systems feature similar hardware and software architectures, Frontier offers more compute nodes, enabling superior scalability for large-scale computations. Each compute node in Defiant is equipped with 64-core AMD EPYC 7662 “Rome” CPUs and 256 GB of memory. Frontier nodes, on the other hand, feature 64-core AMD “Optimized 3rd Gen EPYC” CPUs and provide 512 GB of memory per node.

For efficient task allocation, Defiant is employed for smaller problems requiring fewer than 30 compute nodes (1,500 CPU cores), benefiting from shorter queue times. Conversely, Frontier is reserved for larger problems that demand over 1,500 CPU cores (30 compute nodes), leveraging its extensive computational resources for scalability. All distributed computations, including the execution of distributed variational quantum optimization algorithm (DVQOA), distributed deep neural network (DDNN), and brute-force searches, are performed using a message-passing interface (MPI) implementation.

\subsection*{Quadratic unconstrained binary optimization (QUBO) problems}

QUBOs model the energy function of a certain system \cite{zaman2021pyqubo}, and real-world problems are generally represented as fully connected QUBOs \cite{kim2024distributed}, where all variables interact with one another. To simulate such complex real-world scenarios, QUBO instances used in this study are constructed as fully connected matrices with random elements uniformly distributed between -1 and 1. The energy of the QUBO serves as a cost function for optimization.

Brute force search guarantees the identification of the ground truth for QUBO problems \cite{giuffrida2022engineering}, with an exponential complexity of \textit{O}(2$^n$). For problem size \textit{n}~$<$~40, the ground truths are determined by brute force search with HPC systems, leveraging up to 8,192 CPU cores (164 compute nodes). These solutions serve as references for calculating the approximation ratio, which is defined as: Approximation Ratio = (Identified Value / Ground Truth). The total brute force computation time is estimated by multiplying the runtime of an MPI task with the number of cores employed, i.e., Brute Force Time = MPI Task Runtime~$\times$~Number of Cores. This represents the total time required if brute force is performed on a single core. The generated QUBO instances are further utilized for hyperparameter studies, exploring the effects of hyperparameters (\textit{m} and \textit{t}).

\subsection*{Max-Cut problems}

Max-Cut problems are a fundamental class of binary combinatorial optimization problems, where each variable can take binary values (0 or 1), and the objective considers self- and pairwise interactions between variables \cite{commander2009maximum}. As the number of nodes (\textit{n}) increases, the potential number of edges grows significantly, with a maximum of \textit{n}(\textit{n}-1)/2. To adjust problem complexity, the sparsity of Max-Cut instances can be controlled by varying the number of edges. For this study, we generate relatively dense configurations with the number of edges set to \textit{n}(\textit{n}-1)/8. 

Numerous classical and quantum solvers have been developed to solve Max-Cut problems, with hybrid quantum annealing (HQA; D-Wave Systems, Advantage system 4.1) recognized as one of the most effective solvers \cite{irie2021hybrid}. As demonstrated in Table S1, HQA achieves solutions close to or slightly better than the best-known values, though occasional suboptimal results are also observed \cite{schuetz2022combinatorial, benlic2013breakout, kochenberger2013solving}. Brute force search, while guaranteeing exact solutions, becomes infeasible for large problem sizes (\textit{n} $>$ 40) due to its exponential time complexity. Consequently, for such large Max-Cut problems, the solutions obtained from HQA are used as reference values to calculate the approximation ratio, defined as: Approximation Ratio = (Identified Value / Reference Value). This approach ensures a reliable evaluation of algorithm performance across both moderate and large-scale Max-Cut problem instances. As Max-Cut problems can be formulated as QUBO problems, the QUBO energy serves as the cost function for optimization.

\subsection*{Traveling salesman problems}

The traveling salesman problem (TSP) is another example of a combinatorial optimization problem where quantum computing has the potential to exhibit a \revnewtwo{practical benefit}. TSP is classically challenging due to the factorial growth of possible routes (\textit{n}!), where \textit{n} is the number of cities to visit. In our experiments, cities are randomly distributed within the range [0,10], and the objective is to minimize the total travel distance required to visit every city exactly once. To encode these problems as QUBOs not to visit the same cities, penalty factors are applied with a value of 100. The ground truth for the cost function, representing the shortest path, is determined using brute-force search.

DVQOA is employed to solve the QUBOs that represent the TSP, with the obtained solutions converted into city indices representing the optimal order of cities to visit. Simulated annealing (SA), a classical solver that search optimization spaces by modeling thermal fluctuation, is used to solve the TSP, but it usually fails to find the optimal routes. HQA, recognized as one of the best solvers for QUBO-type problems as demonstrated in Max-Cut problems, is also applied to the TSP. Solution quality is evaluated using the approximation ratio, defined as Approximation Ratio = (Cost$_{global}$ / Cost$_{solver}$). By definition, an approximation ratio close to 1 indicates a high-quality solution.

Hyperparameters in DVQOA are adjusted to solve more complex problems: m and t are set to 3 for problems with fewer than 7 cities, and 7 for problems with 7 and 9 cities.

\subsection*{Chemistry problems}

Quantum-classical algorithms have been widely explored for quantum chemistry calculations, particularly for determining the lowest energy states of molecules. Since the computational cost scales exponentially with a system size, variational quantum eigensolvers (VQEs) have gained significant attention as a promising approach leveraging quantum principles \cite{xu2024truncation, tran2024variational}.

In this study, we select eight molecules—Hydrogen (H\textsubscript{2}), Hydrogen Fluoride (HF), Lithium Hydride (LiH), Hydrogen Dioxide (H\textsubscript{2}O\textsubscript{2}), Beryllium Hydride (BeH\textsubscript{2}), Ammonia (NH\textsubscript{3}), Methane (CH\textsubscript{4}), and Acetylene (C\textsubscript{2}H\textsubscript{2})—to compute their ground-state energies. The molecular Hamiltonian is mapped onto a qubit operator to construct the cost Hamiltonian using qiskit-nature and pyscf. A two-local ansatz, composed of Ry gates and CNOT gates, is employed for energy estimation using VQE, implemented in qiskit-algorithms. The COBYLA optimizer is used for classical optimization, with a maximum of 500 iterations. Parameters are initialized within the range [-2$\pi$, 2$\pi$], and energy values are estimated using a noiseless simulator (AerEstimator). VQE is executed in a distributed computing environment with 500 cores across 10 compute nodes.

Our DVQOA efficiently calculates eigenvalues for quantum chemistry problems using a state-vector estimator, eliminating the need for computational state measurements. This approach allows the eigenvalues to be determined directly from circuit parameters and the cost Hamiltonian. The lowest molecular energy (ground state) is obtained using the 'NumPyMinimumEigensolver' on a classical computer, serving as a reference for computing the approximation ratio: Approximation Ratio = (Energy$_{solver}$ / Ground State).

\subsection*{Layered photonic structures}

Layered photonic structures serve as a practical testbed to show the efficiency of DVQOA in solving complex optimization problems within material science. These structures can have unique optical characteristics, such as transmitting visible photons while reflecting ultraviolet (UV) and near-infrared (NIR) photons, making them suitable for optical filter applications like energy-saving windows \cite{kim2022high, kim2024wide, xu2024quantum}. Furthermore, by incorporating a thermal radiative layer on the top surface, such photonic structures can have radiative cooling performance by emitting thermal radiation through an atmospheric window (8 $\mu$m $< \lambda <$ 13 $\mu$m) to mitigate the global warming issue \cite{li2023ultrathin}. Hence, these structures can be employed for developing transparent radiative coolers \cite{kim2021visibly, kim2022high, kim2024wide}. 

Fig. S11 illustrates the design involves a silica substrate with a top layer of 40 $\mu$m-thick polydimethylsiloxane (PDMS), and the thickness of the photonic structure is set to 1.2 $\mu$m. Each layer in the photonic structures can be one of four materials: silicon dioxide (SiO$_2$), silicon nitride (Si$_3$N$_4$), aluminum oxide (Al$_2$O$_3$), and titanium dioxide (TiO$_2$). They can be encoded with two-digit binary labels: ‘00’ for SiO$_2$, ‘01’ for Si$_3$N$_4$, ‘10’ for Al$_2$O$_3$, and ‘11’ for TiO$_2$. Hence, a 6-layered photonic structure corresponds to a bitstring length of 12 (i.e., problem size or circuit width \textit{n} = 12). Photons can be transmitted or reflected depending on the refractive contrast at the interface between layers, thus optimizing layer configuration is important to achieve desirable optical characteristics. However, an exponentially large design space arising from 2$^{n}$ possible configurations makes it challenging to solve these optimization problems.

Energy-saving windows aim to maximize transmitted irradiance in the visible range while minimizing it in the UV/NIR ranges \cite{kim2022high, kim2021visibly}. An ideal design achieves perfect transmission efficiency in the visible range, and zero transmission in other ranges (Fig. S13). To evaluate the performance of a designed photonic structure, a figure-of-merit (FOM) is introduced \cite{kim2022high, kim2024wide}:

\begin{equation}
        TI(\lambda) = T(\lambda) \times S(\lambda),
        \label{eq:sup_example1} 
\end{equation}

\begin{equation}
        \text{FOM} = \frac{10 \int_{300}^{2500} \left(TI_{\text{designed}}(\lambda) - TI_{\text{ideal}}(\lambda)\right)^2 \, d\lambda}{\int_{300}^{2500} S(\lambda)^2 \, d\lambda},
        \label{eq:sup_example2} 
\end{equation}

\noindent
where \textit{T}($\lambda$) and \textit{S}($\lambda$) respectively represent transmission efficiency and the solar spectral irradiance (under air mass 1.5 global \cite{li2022tandem}). Here, transmission efficiency \textit{T}($\lambda$) is calculated with the transfer matrix method, a computationally efficient methodology for layered structures \cite{luce2022tmm}. \textit{TI}($\lambda$) is the transmitted irradiance through the designed or ideal photonic structures. By definition, a lower FOM indicates that the optical properties of the designed structure closely match those of the ideal structure, indicating a superior design. Consequently, this problem aligns well with the minimization capabilities of DVQOA, which effectively uses FOM as the cost function, making it an effective algorithm for addressing such complex material design challenges.

\subsection*{Metamaterial optical diodes}

Optical diodes are essential components in photonics, designed to allow unidirectional light transmission while blocking light in the reverse direction \cite{kim2024quantum, shen2015integrated, tang2016broadband}. They are crucial for protecting optical sources and processing optical information efficiently \cite{kim2024quantum}. Thin optical diode designs often leverage pixelated metamaterials, where each pixel is assigned as either dielectric (‘0’) or metallic (‘1’) materials (Fig. S12). These metamaterial optical diodes should achieve high forward transmission ($T_F$)and low (near-zero) backward transmission ($T_B$) to exhibit effective unidirectional optical behavior. 

Asymmetric transmission is realized by enabling first-order diffraction only for forward incident light. This behavior is achieved when the following condition is satisfied:

\begin{equation}
        \frac{2n_1\pi}{\lambda_0} < \frac{2\pi}{\Lambda_G} < \frac{2n_2\pi}{\lambda_0},
        \label{eq:sup_example11} 
\end{equation}

\noindent
where \textit{n\textsubscript{1}} and \textit{n\textsubscript{2}} are the refractive indices of the upper and lower media, respectively. $\Lambda_G$ represents the periodic length of the unit cell, and $\lambda_0$ denotes the wavelength of the incident light. In this study, we set the parameters \textit{n\textsubscript{1}}, \textit{n\textsubscript{2}}, $\Lambda_G$, and $\lambda_0$ to 1 (air), 1.45 (silicon dioxide), 600 nm, and 800 nm, respectively. Under these conditions, we optimize pixel configurations within the unit cell of a metamaterial optical diode to achieve the desired unidirectional optical behavior.

To evaluate the performance of a designed optical diode, FOM is defined as: FOM = $T_B$ – $T_F$. Here, transmission efficiency is calculated using rigorous coupled wave analysis, which is a semi-analytical method for solving Maxwell's equation \cite{kim2024meent}. The optimization of pixelated metamaterial configurations is important to achieve the desired optical diode performance, characterized by a low FOM. However, the optimization space grows exponentially (2$^{n}$ possible configurations), with the number of pixels \textit{n}, making this problem computationally challenging. Our DVQOA is particularly well-suited for addressing this type of minimization problem (with FOM as the cost function), offering a promising approach for exploring the large material optimization space efficiently to identify the optimal structure.

\subsection*{Distributed deep neural networks}

Deep neural networks (DNNs) are well-suited for binary optimization tasks, leveraging the sigmoid activation function to output probabilities associated with binary values for each variable. Our DNN architecture consists of two hidden layers with 128 and 64 neurons, respectively. The Adam optimizer is employed with a learning rate of 0.001, and the network is trained over 500 epochs to minimize the cost function. Input data is initialized within the range [0,1]. The cost function corresponds to the energy state of a given problem, enabling the DNN to identify a binary vector that minimizes the cost function of the optimization problem.

To improve performance, the distributed execution of DNNs (DDNN) across multiple cores and nodes is implemented using MPI. By utilizing 500 cores distributed across 10 compute nodes, similar to our distributed quantum approach (DVQOA), DDNN enhances the chance of identifying the ground truth solution through diverse parameter configurations. Among the solutions generated by DDNN, the best solution is selected to calculate the approximation ratio. The time to solution is measured from the task initiation to its completion.


\makeatletter
\setcounter{figure}{0}

\makeatother

\section*{Data availability}
All data supporting the findings of this study are available within the paper and its Supplementary Information. Information requests are available from the corresponding author upon reasonable request. 

\section*{Code availability}
All codes that support the findings of this study are available upon request to the corresponding authors.

\section*{Acknowledgments}
We thank Nils Herrmann and John L. Helm from Quantum Brilliance, and Vincent R. Pascuzzi from IBM for their insightful discussion. This research used resources of the Oak Ridge Leadership Computing Facility at the Oak Ridge National Laboratory, which is supported by the Office of Science of the U.S. Department of Energy under Contract No. DE-AC05-00OR22725. This material is based upon work supported by the U.S. Department of Energy, Office of Science, National Quantum Information Science Research Centers, Quantum Science Center. 

Notice: This manuscript has in part been authored by UT-Battelle, LLC under Contract No. DE-AC05-00OR22725 with the U.S. Department of Energy. The United States Government retains and the publisher, by accepting the article for publication, acknowledges that the U.S. Government retains a non-exclusive, paid up, irrevocable, world-wide license to publish or reproduce the published form of the manuscript, or allow others to do so, for U.S. Government purposes. The Department of Energy will provide public access to these results of federally sponsored research in accordance with the DOE Public Access Plan (http://energy.gov/downloads/doe-publicaccess-plan).

\section*{Author Contributions}
Conceptualization, methodology, investigation, data collection, data analysis and visualization: S.K.; Expertise: S.K., and I.S.; Writing: S.K., and I.S.

\section*{Competing Interests}
The authors declare no competing interests.

\printbibliography
\clearpage
\section*{Figures}
\begin{figure} 
	\centering
	\includegraphics[width=1\textwidth]{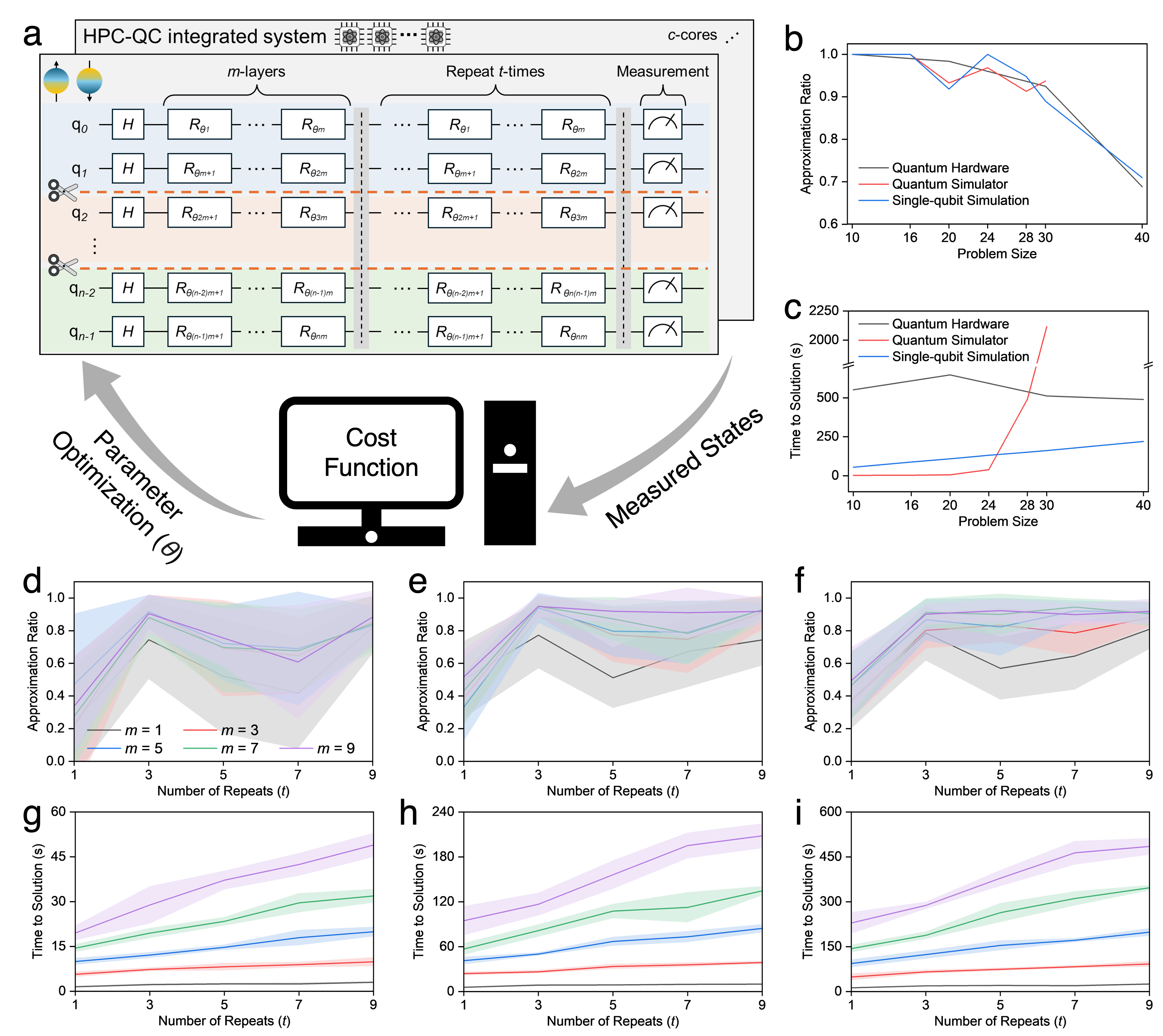} 

    \caption{\textbf{Workflow and performance of VQOA.}
	\textbf{a}, Schematic representation of VQOA workflow, featuring the efficient ansatz and the iterative adjustment of rotation gate parameters through classical optimization to minimize the cost function. The problem size corresponds to circuit width (\textit{n}), and HPC-QC integrated systems enable distributed executions of VQOAs (DVQOA). 
    Approximation ratio (\textbf{b}) and 
    time to solution (\textbf{c}) for VQOA on quantum hardware (\texttt{IBM-Quantum-Device}) and quantum simulator (\texttt{Qiskit-Aer}). Here, each qubit is simulated independently for single-qubit simulations. 
    While quantum hardware maintains \revnewtwo{nearly} constant time complexity, the simulator exhibits exponential growth in time, whereas single-qubit simulations scale linearly. 
    \textbf{d -- i}, Hyperparameter study: the number of layers \textit{m} and repeats \textit{t}. Approximation radio (\textbf{d},\textbf{e},\textbf{f}) and time to solution (\textbf{g},\textbf{h},\textbf{i}) with different hyperparameters. For this study, \textit{n} = 10 (\textbf{d},\textbf{g}), \textit{n} = 20 (\textbf{e},\textbf{h}), and \textit{n} = 30 (\textbf{f},\textbf{i}). 
    By implementing repeated gates while keeping the number of trainable parameters smaller, and by eliminating entangling gates while enabling parameters to naturally learn complex correlations, the Barren Plateaus problem is effectively mitigated, leading to high approximation ratios.  
     }
	\label{fig:1} 
\end{figure}

\clearpage
\begin{figure} 
	\centering
	\includegraphics[width=1\textwidth]{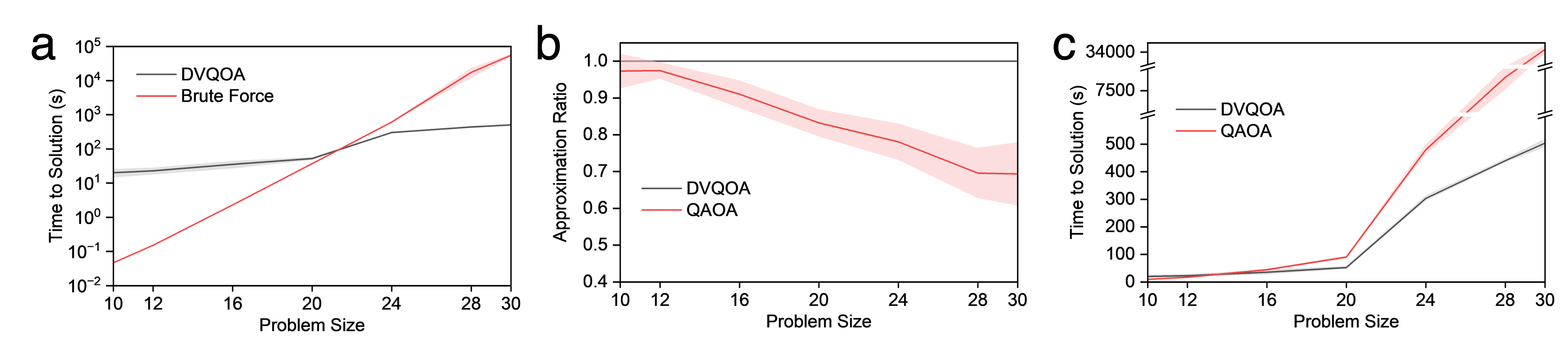} 

    \caption{\textbf{Scalability and efficiency of DVQOA.}
	\textbf{a}, Comparison of the time to solution between DVQOA and brute force. Brute force requires exponential time \textit{O}(2$^{n}$) to identify the ground truth as \textit{n} increases, while DVQOA exhibits a linear algorithmic complexity. 
    Approximation ratio (\textbf{b}) and time to solution (\textbf{c}) of DVQOA and QAOA for QUBO problems as a function of \textit{n}. While QAOA shows degraded performance with increasing \textit{n}, DVQOA maintains stable performance across varying \textit{n}.   }
	\label{fig:2} 
\end{figure}

\clearpage
\begin{figure} 
	\centering
	\includegraphics[width=1\textwidth]{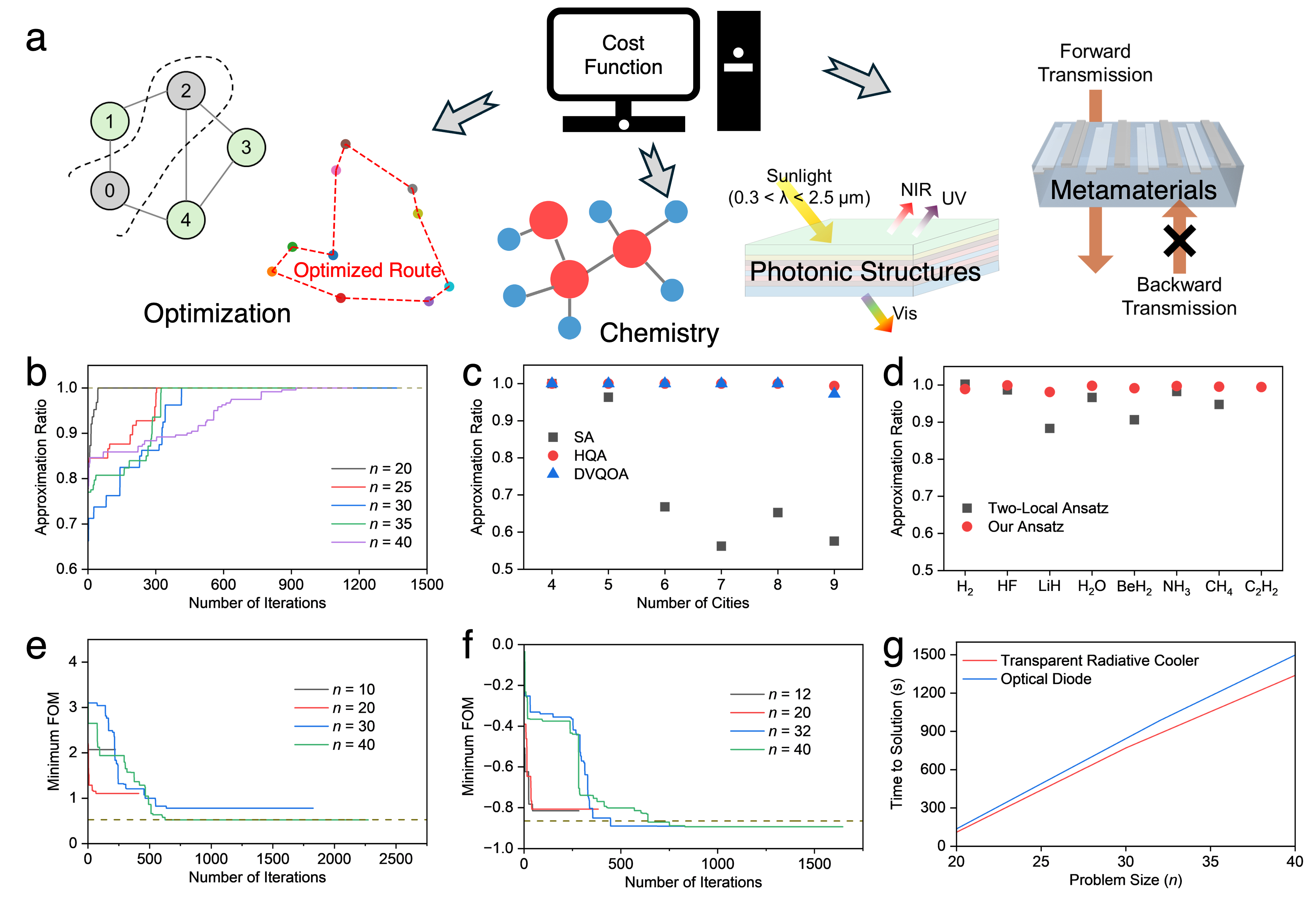} 

	\caption{\textbf{Applications of DVQOA to solve various problems.}
        \textbf{a}, Schematic representation of DVQOA’s versatility, illustrating its ability to address various real-world problems by redefining the cost function. 
        \textbf{b}, Evolution of the approximation ratio for Max-Cut problems with various \textit{n}. 
        \textbf{c}, Approximation ratio for TSP with varying number of cities to visit.
        \textbf{d}, Approximation ratio for computing the minimum eigenvalue of molecular Hamiltonians, highlighting DVQOA’s capability as an eigensolver.
        \textbf{e, f}, Evolution of FOM for real-world material optimization challenges: layered photonic structures (\textbf{e}) and metamaterial optical diodes (\textbf{f}). Dotted lines indicate the best-known results, highlighting DVQOA's superior performance in identifying high-quality solutions. These optimization examples exhibit an exponential optimization space (2$^{n}$). \textbf{g}, Time to solution for completing material optimization tasks as a function of \textit{n}, demonstrating DVQOA's capability for completing such complex optimization tasks with a linear complexity \textit{O}(\textit{nmt}). }
	\label{fig:3} 
\end{figure}

\clearpage
\begin{figure} 
	\centering
	\includegraphics[width=1\textwidth]{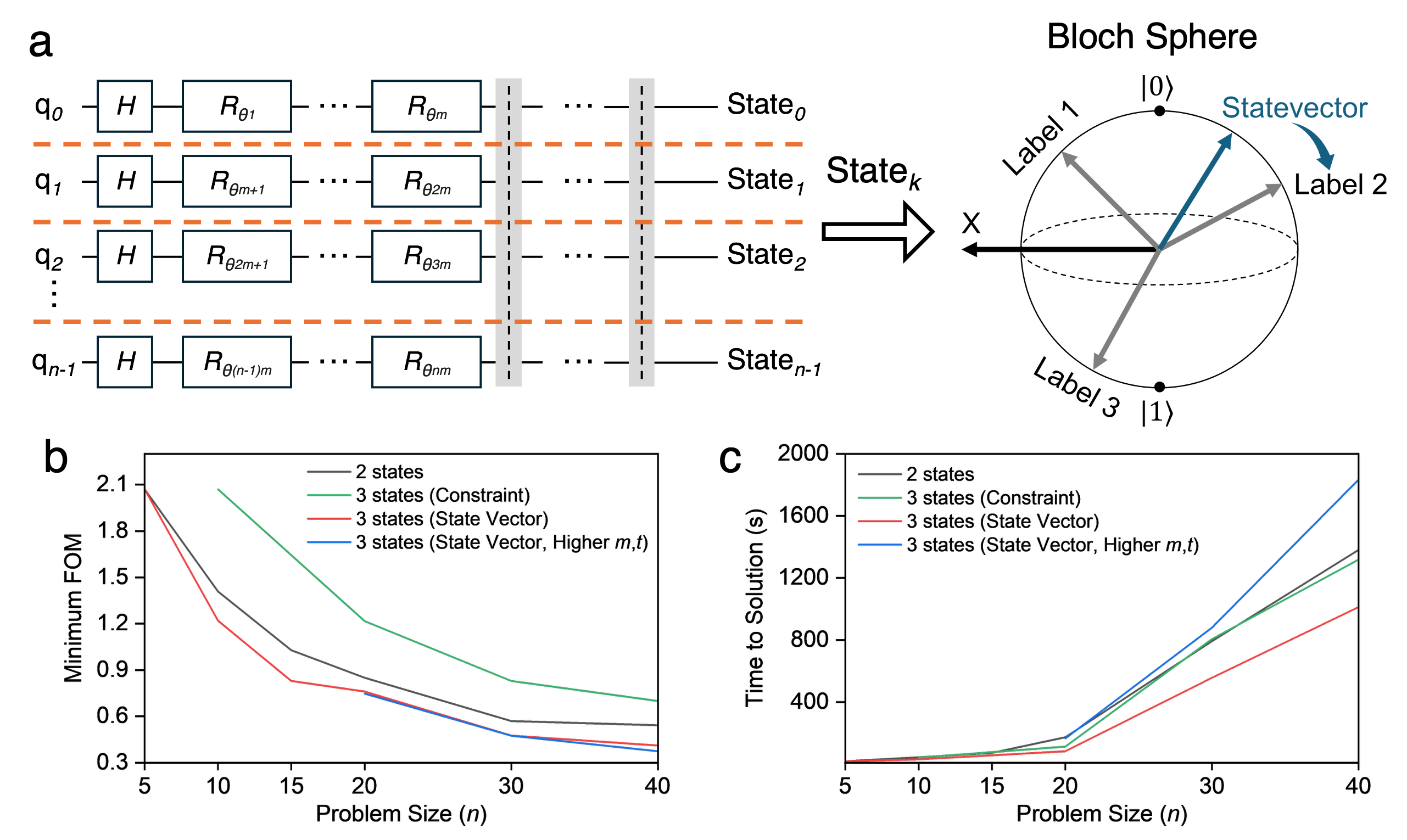} 

	\caption{\textbf{State vector on the Bloch sphere for \textit{N}-ary optimization with DVQOA.}
	\textbf{a}, Schematic illustrating how each state is encoded as a unique vector on the Bloch sphere, allowing a single qubit to represent one of \textit{N} possible states without requiring additional qubits or constraints. \textbf{b}, Minimum FOM as a function of \textit{n} for material optimization tasks. Four DVQOA methods are compared: (i) two-state encoding per qubit, (ii) three-state encoding with additional qubits and constraints, (iii) three-state encoding using a state vector on the Bloch sphere, and (iv) three-state encoding using a state vector on the Bloch sphere with increased \textit{m} and \textit{t}. \textbf{c}, Time to solution for completing optimization tasks as a function of \textit{n}. }
	\label{fig:4} 
\end{figure}


\clearpage
\newpage

\begin{table}[t] 
	\centering
        \scriptsize
	\caption{\textbf{Performance of DVQOA for higher-order (\textit{k} $\geq$ 3) binary optimization problems (\textit{N} = 2).}
		Approximation ratio and time to solution achieved by DVQOA and DDNN for different interaction orders (\textit{k}) and problem sizes (\textit{n}). Higher-order problems include all interactions of lower orders; e.g., 3$^{rd}$-order problems involve self-, pairwise, and three-variable interactions, significantly increasing problem complexity. Problem sizes range from 24 to 30, with instances labeled 'P\textit{k}-\textit{n}'. The results demonstrate that DVQOA identifies the ground truth with substantially shorter times for complex problems, compared to brute force search and DDNN, emphasizing the practical \revnewtwo{benefit} of utilizing the quantum algorithm. }
	\label{tab:1} 
        \setlength{\tabcolsep}{2pt}
	\begin{tabular}{lccccccccc}
            \\
            \hline
            \textbf{\textit{k}} & \multicolumn{3}{c}{\textbf{3}} & \multicolumn{3}{c}{\textbf{4}} & \multicolumn{3}{c}{\textbf{5}} \\  \hline
            \textbf{Instances} & \textbf{P3-26} & \textbf{P3-28} & \textbf{P3-30} & \textbf{P4-24} & \textbf{P4-26} & \textbf{P4-28} & \textbf{P5-22} & \textbf{P5-24} & \textbf{P5-26} \\ \hline
            \textbf{DVQOA Approximation Ratio} & 1 & 1 & 1 & 1 & 1 & 1 & 1 & 1 & 1\\
            \textbf{DVQOA Time to Solution (s)} & 392.685 & 469.775 & 530.158 & 369.161
            & 402.766 & 498.876 & 332.853 & 487.724 & 596.293 \\
            \textbf{DDNN Approximation Ratio} & 0.9787 & 1 & 1 & 0.9903
             & 1 & 1 & 1 & 1 & 1 \\
            \textbf{DDNN Time to Solution (s)} & 328.226   &398.4923  &497.4745  &1850.603  &2573.306  &3534.801  &6269.772  &10164.69  &15549.36  \\
            \textbf{Brute Force Time (s)} & 1,213,031 & 3,581,267 & 18,239,898 & 1,416,661
            & 7,625,766 & 25,148,563 & 1,011,141 & 4,345,634 & 25,107,394 \\
            \textbf{Acceleration} & 0.8358  &  0.8482  &0.9383  &5.0129  &6.3890  &7.0855  &18.8364  &20.8410  &26.0766 \\
            \hline
 
        \end{tabular}  
\end{table}

\clearpage
\newpage

\begin{table} 
	\centering
        \small
	\caption{\textbf{Performance of DVQOA for \textit{N}-ary problems involving higher-order interactions.}
		Approximation ratio and time to solution of DVQOA for \textit{N}-ary problems (\textit{N} = 2, 3, 4, and 5) involving 2$^{nd}$-order and 3$^{rd}$-order interactions, for \textit{n} = 12. The table demonstrates that while brute force search requires exponential time to identify the ground truth, DVQOA achieves the ground truth effectively. Here, the hyperparameters (\textit{m}, \textit{t}) are both 3 for \textit{N} $\leq$ 4, and are both 7 for \textit{N} = 5.}
	\label{tab:2} 
	
	\begin{tabular}{lcccccccc} 
            \\
		\noalign{\global\arrayrulewidth=0.5mm}
            \hline
            \textbf{\textit{k}} & \multicolumn{4}{c}{\textbf{2}} & \multicolumn{4}{c}{\textbf{3}} \\ \hline
            \textbf{\textit{N}} & \textbf{2} & \textbf{3} & \textbf{4} & \textbf{5} & \textbf{2} & \textbf{3} & \textbf{4} & \textbf{5} \\ \hline
            \textbf{Approximation Ratio} & 1 & 1 & 1 & 1 & 1 & 1 & 1 & 1 \\ 
            \textbf{Time to Solution (s)} & 6.083 & 6.112 & 6.986 & 38.42 & 5.184 & 6.486 & 9.654 & 36.10 \\ 
            \textbf{Brute Force Time (s)} & 2.281 & 121.7 & 10,739 & 127,990 & 8.639 & 485.5 & 34,134 & 471,200 \\ 
             \hline 

	\end{tabular}
\end{table}

\includepdf[pages=-]{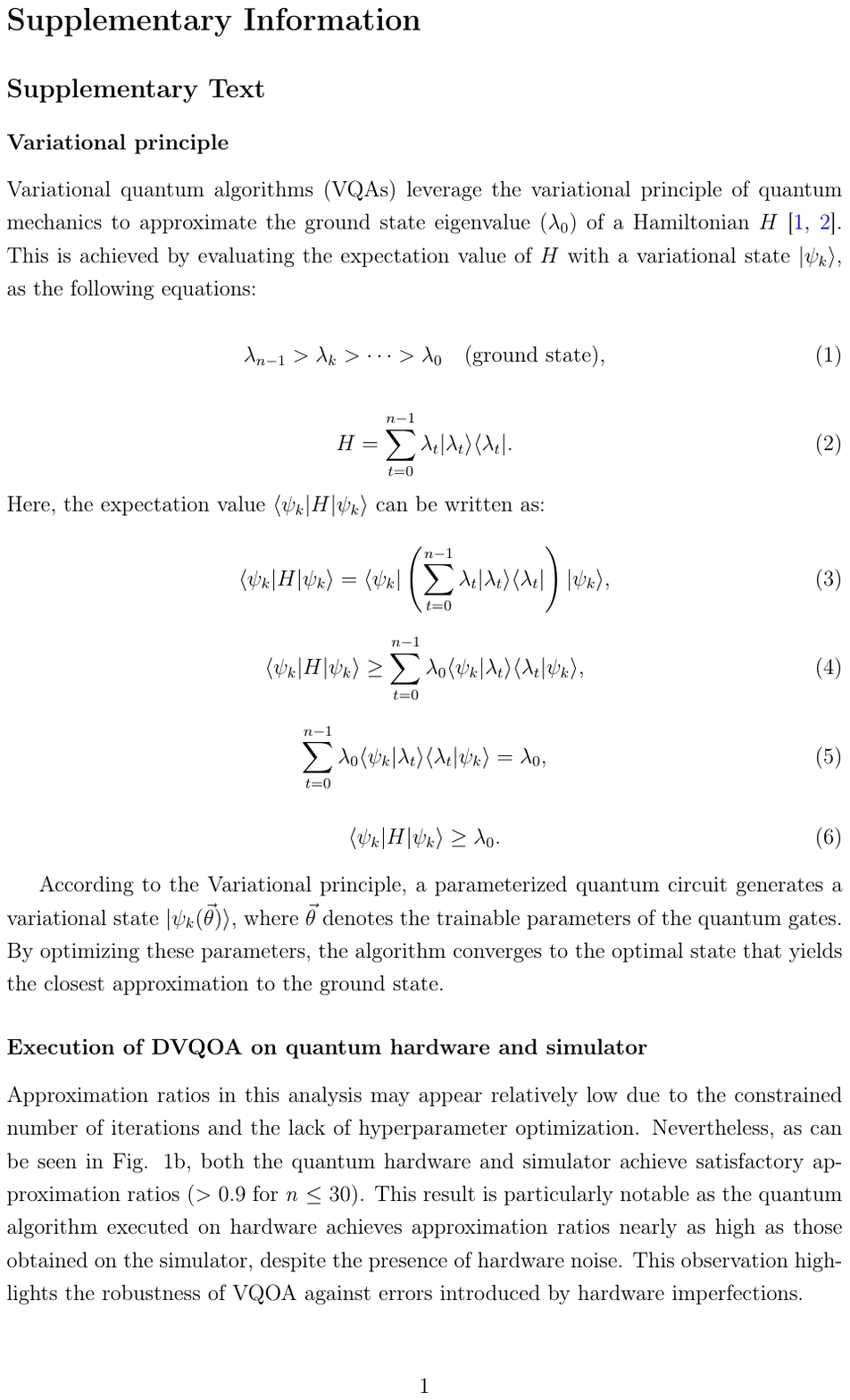}

\end{document}